\documentclass[aps,prb,reprint,superscriptaddress]{revtex4-1}
\usepackage{graphicx} 
\usepackage{amsmath,amsfonts,amssymb}

\begin{document}

% Use the \preprint command to place your local institutional report
% number in the upper righthand corner of the title page in preprint mode.
% Multiple \preprint commands are allowed.
% Use the 'preprintnumbers' class option to override journal defaults
% to display numbers if necessary
%\preprint{}

%Title of paper
\title{Synchronization of electrically coupled stochastic magnetic oscillators induced by thermal and electrical noise}

\author{A. Mizrahi}
\affiliation{Unite Mixte de Physique CNRS / Thales, Univ. Paris-Sud, Universite Paris-Saclay, 91767 Palaiseau, France}
\affiliation{Institut d’Electronique Fondamentale, Univ. Paris-Sud, CNRS, Universite Paris-Saclay, 91405 Orsay, France}
\author{N. Locatelli}
\affiliation{Institut d’Electronique Fondamentale, Univ. Paris-Sud, CNRS, Universite Paris-Saclay, 91405 Orsay, France}
\author{J. Grollier}
\affiliation{Unite Mixte de Physique CNRS / Thales, Univ. Paris-Sud, Universite Paris-Saclay, 91767 Palaiseau, France}
\author{D. Querlioz}
\affiliation{Institut d’Electronique Fondamentale, Univ. Paris-Sud, CNRS, Universite Paris-Saclay, 91405 Orsay, France}

\date{\today}

\begin{abstract}
%damien: j ai pas mal retravaille l abstract
Superparamagnetic tunnel junctions are nanostructures that  auto-oscillate stochastically under the effect of thermal noise.
Recent works showed that despite their stochasticity, such junctions possess a capability to synchronize to  subthreshold  voltage drives, in a way that can be enhanced or controlled by adding noise.
In this work, we investigate a system composed of two electrically coupled junctions, connected in series to a periodic voltage source. 
We make use of numerical simulations and of an analytical model to demonstrate that both junctions can be phase-locked to the drive, in phase or in anti-phase. 
This synchronization phenomenon can be controlled by both thermal and electrical noises, 
although the two types of noises induce qualitatively different behaviors. 
Namely, thermal noise can stabilize a regime where one junction is phase-locked to the drive voltage while the other is blocked in one state. 
On the contrary, electrical noise causes the junctions to have highly correlated behaviors and thus cannot induce the latter.  
%damien: rajouter une phrase de perspective. demander a julie si c est trop applique
These results open the way for the design of superparamagnetic tunnel junctions that can perform computation through synchronization, and which  harvest  the largest part of  their energy consumption  from thermal noise.
\end{abstract}


\maketitle

\section{Introduction}
Superparamagnetic tunnel junctions are stochastic auto-oscillators, powered by thermal noise~\cite{rippard_thermal_2011}. They naturally transform the fluctuations of thermal noise into a large telegraphic signal that can be read through a simple measurement of their electrical resistance. Furthermore, they have the ability to exhibit the phenomenon of stochastic resonance in which noise enables a weak signal to be detected by a non linear system~\cite{cheng_nonadiabatic_2010,finocchio_micromagnetic_2011,daquino_stochastic_2011}. In previous works, we showed how they can also synchronize to a weak periodic voltage through both thermal noise and electrical noise on the drive~\cite{locatelli_noise-enhanced_2014,mizrahi_controlling_2016}. The capacity of superparamagnetic tunnel junctions to harvest the energy of noise gives the opportunity to study their physics under many angles, from sensing to low-power computing.
%makes them promising candidates for low-power applications of all kinds, from sensing to computing. 
This requires understanding how these noise-induced phenomena occur in systems composed of several coupled superparamagnetic tunnel junctions. Noise-induced synchronization of coupled elements has been studied for several systems, such as arrays of generic nonlinear elements~\cite{neiman_stochastic_1995,lindner_array_1995, marchesoni_spatiotemporal_1996} or linear chains of diffusely coupled diode resonators~\cite{locher_stochastic_1998}, but their analysis does not apply to our specific device as they consider linear coupling and collective phenomena. Furthermore, these works focus on noise-induced signal power amplification  but do not study frequency-locking and phase-locking directly. 
 
In this work, we present a theoretical study of noise-induced synchronization of two electrically coupled superparamagnetic tunnel junctions. Our comprehensive study is based on numerical simulations. We examine frequency-locking through the evolution with noise of the junctions frequencies as well as phase-locking through time-resolved traces of the junctions resistances. All results are discussed with the support of an analytical model based on the evolution with noise of the junctions switching probabilities. Both numerical and analytical models were previously validated by direct comparison to experiments in the case of the synchronization of a single junction~\cite{mizrahi_magnetic_2015,mizrahi_controlling_2016}. We examine the differences between the synchronization phenomena induced by thermal noise versus those induced by electrical noise on the drive voltage.

After presenting our model in the case of a single superparamagnetic tunnel junction, we apply it to two junctions connected in series to a periodic voltage source. We show that both junctions can be synchronized simultaneously to the voltage drive. Both in-phase-locking and anti-phase-locking configurations can be achieved by choosing the connections of
the junctions in the circuit. These phenomena can be induced and controlled by the amplitudes of thermal and electrical noises. Specifically, we show that thermal noise is able to stabilize a hybrid synchronization regime where at each
oscillation of the drive, one randomly chosen junction is phase-locked to the drive while the other is blocked in one state. On the other hand, this regime cannot be achieved by the addition of electrical noise. The electrical noises on the two junctions both come from the drive and are thus correlated. The switches of the junctions are simultaneous and triggered by large voltage events.

\section{Noise induced synchronization of a single junction}
\begin{figure*}
\includegraphics[scale=0.6]{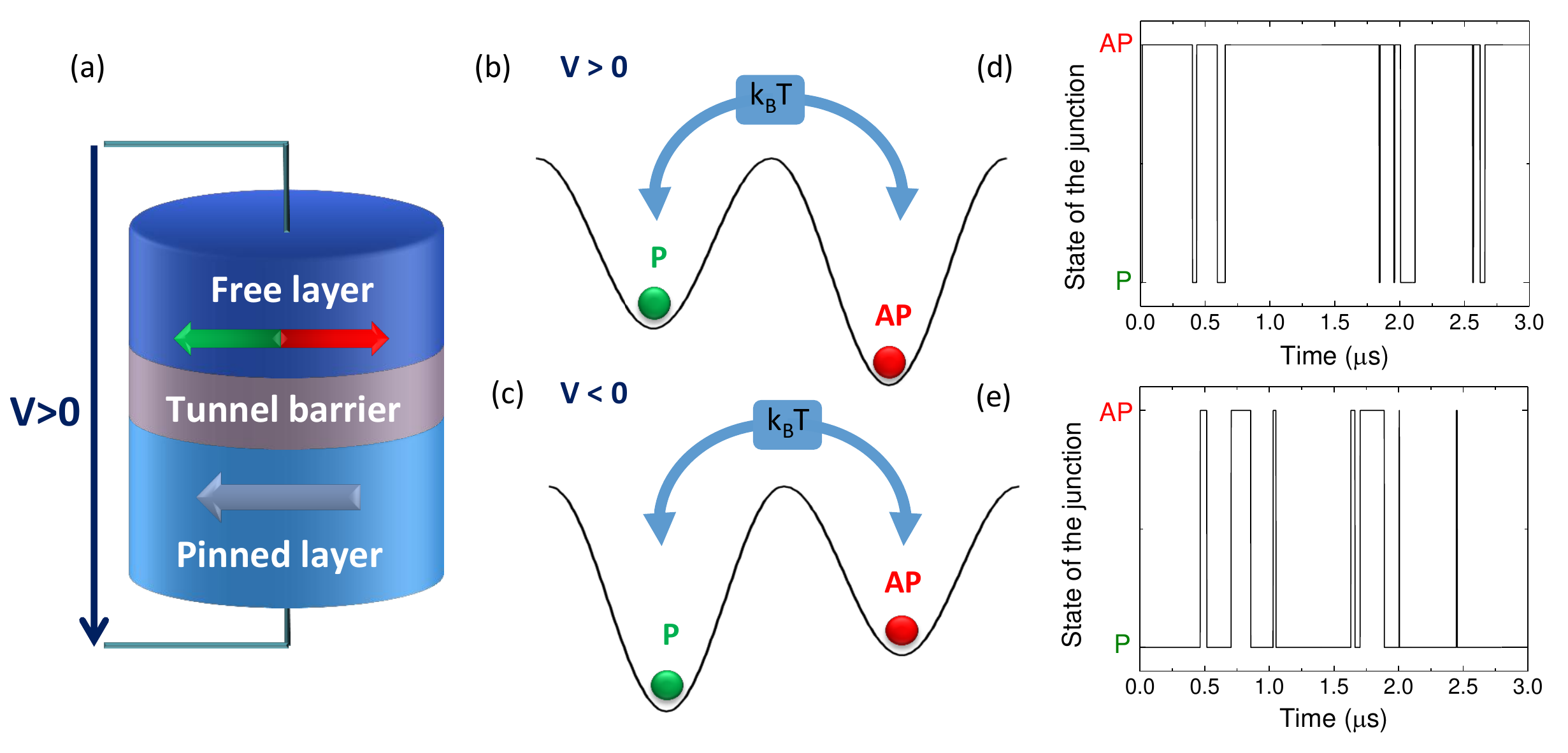}%
\caption{(a) Schematic view of a magnetic tunnel junction. 
(b-c) Energy landscape of the system when a positive (b) or negative (c) voltage is applied. 
(d-e) Numerical simulations of the state of a junction versus time for applied voltages of $V=+0.17 V_c$ and $V=-0.17 V_c$, at a temperature $T=300\mathrm{K}$. 
\label{junction}}
\end{figure*}

A magnetic tunnel junction is composed of two ferromagnets (one with pinned magnetization and one with free magnetization) and a tunnel barrier sandwiched between them (Fig.\ref{junction}(a)). The magnetization of the free layer has two stable states: parallel (P) and anti-parallel (AP) to the magnetization of the pinned layer. The anti-parallel state has a higher electrical resistance than the parallel state. Magnetic tunnel junctions can be engineered so that the energy barrier between the P and AP states is low enough to reach the superparamagnetic regime. This means that the intrinsic thermal noise in the junction induces random, spontaneous switches of the magnetization between the P and AP states. The superparamagnetic tunnel junction behaves as a noise-powered stochastic oscillator~\cite{rippard_thermal_2011, locatelli_noise-enhanced_2014} -- as depicted in Fig.\ref{junction}(d-e) -- which does not need any external supply of energy to run. The phenomenon of spin-transfer torque~\cite{slonczewski_current-driven_1996,berger_emission_1996} provides a handle over the junction as a positive voltage stabilizes the AP state (Fig.\ref{junction}(b) and (d)) while a negative voltage stabilizes the P state (Fig.\ref{junction}(c) and (e)). The applied dc voltage is defined as positive from the pinned layer to the free layer. Thermal energy destabilizes both states.\\

\subsection{Methods}
Making the assumption that the free layer of the junction can be considered as a single domain magnetization element, the probabilities to leave the AP and P states within a duration $t$ can be described by a Poisson process:
\begin{align}
P_{AP\rightarrow P}(V,t) &= 1-\exp\left(-\phi_{AP}(V)~t\right),\\
P_{P\rightarrow AP}(V,t) &= 1-\exp\left(-\phi_{P}(V)~t\right).
\label{eq_PP}
\end{align}

The escape rates $\phi_{AP(P)}$ are described by the Neel-Brown model~\cite{brown_thermal_1963}, and follow the Arrhenius equations. They can be tuned by the voltage applied to the junction, through the effect of spin-transfer torque~\cite{slonczewski_excitation_1999,li_thermally_2004}:
\begin{align}
\phi_{AP}(V) &= \phi_0~\exp\left(-\frac{\Delta E}{k_BT}\left(1+\frac{V}{V_c}\right)\right),\\
\phi_{P}(V) &= \phi_0~\exp\left(-\frac{\Delta E}{k_BT}\left(1-\frac{V}{V_c}\right)\right),
\label{eq_tauP}
\end{align}
where $\phi_0=10^9\,\mathrm{Hz}$ 
%damien: mettre environ egal plutot que egal?
is the effective attempt rate~\cite{rippard_thermal_2011}, $\Delta E$ is the energy barrier between the two stable states, $k_B$ is the Boltzmann constant, $T$ is the temperature, $V$ is the voltage across the junctions and $V_c$ is the threshold voltage for deterministic switching at 0K~\cite{li_thermally_2004}. \\

We consider in the following of this article junctions with an energy barrier and temperature such that $\Delta E=30\,k_B T$ at $T=50\mathrm{K}$ and a threshold voltage $V_c=1\,\mathrm{V}$. First, a junction is submitted to a square periodic voltage with frequency $F_{ac}=1/T_{ac}=1\,\mathrm{MHz}$ and with subthreshold amplitude $V~=~0.75\mathrm{V}$.

We perform numerical simulations of the system as follows. At each time step of $dt=10\mathrm{ns}$, the probability for the junction to switch from one state to the other is computed through equations (1) to (4). A uniform pseudo random number between 0 and 1 is generated. If the random number is lower than the switching probability, the junction changes state.

\subsection{Thermal noise induced synchronization \label{syncT}}
 
Synchronization of the junction to the weak periodic signal can be controlled by different sources of noise. We first study the influence of thermal noise. 

\begin{figure*}  
 \includegraphics[scale=0.5]{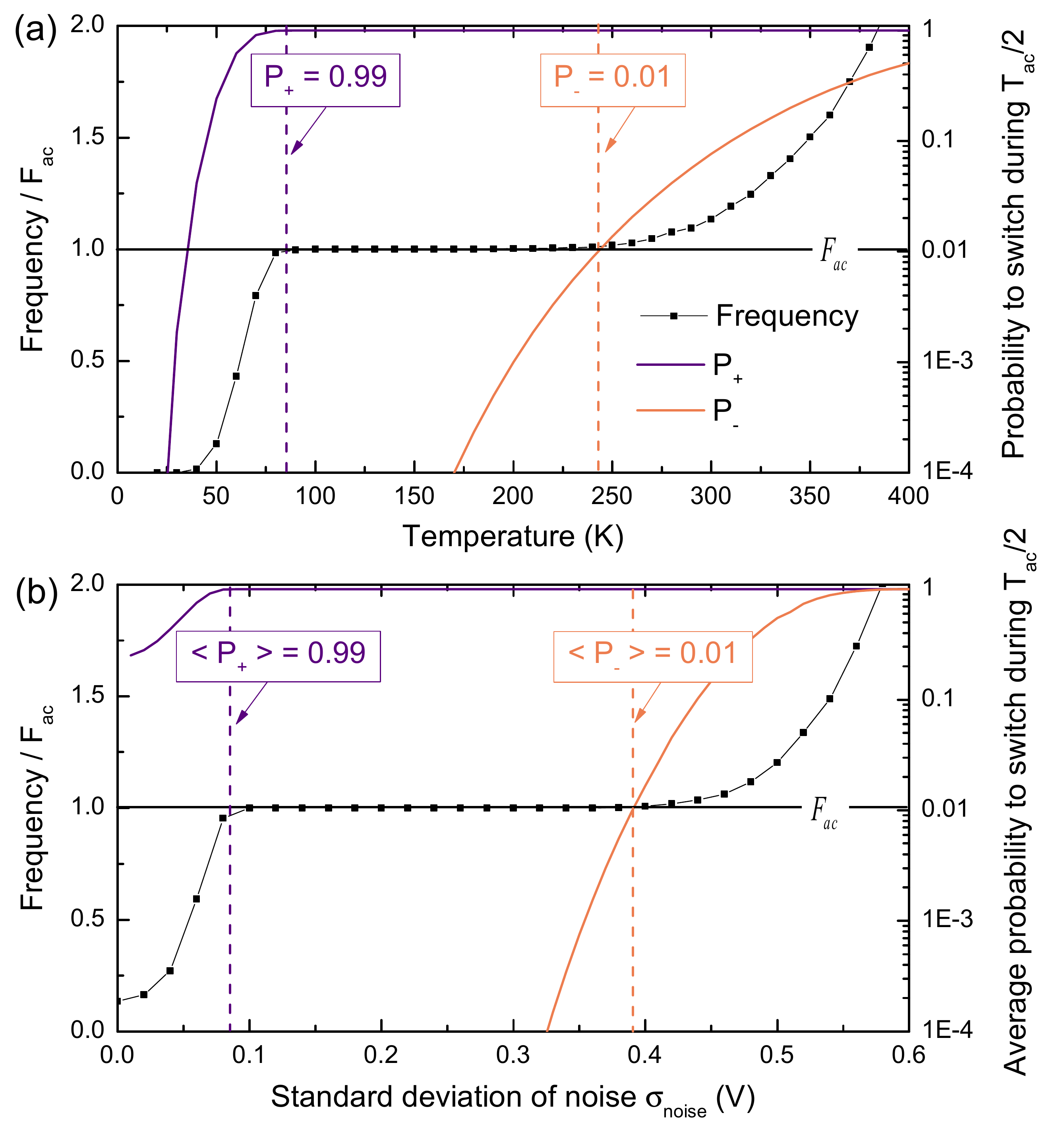}
 \caption{(a) The frequency of the stochastic oscillator is plotted versus temperature in blue squares (in units of the drive frequency $F_{ac}$). The probabilities to phase-lock $P_+$ and unlock $P_-$ are plotted versus temperature in purple and orange solid lines respectively. The drive frequency $F_{ac}$ is represented by a black horizontal line. The temperatures for which $P_+ = 0.99$ and $P_- = 0.01$ are represented by purple and orange vertical dashed lines respectively. 
(b) The frequency of the stochastic oscillator, as well as the average probabilities to phase-lock and unlock, are plotted versus the standard deviation of the white Gaussian electrical noise distribution. The same legend is used for both sub-figures (a) and (b). 
\label{OneSync}}
\end{figure*}

 Fig.~\ref{OneSync}(a) represents the frequency of the junction -- defined as the mean number of oscillations of the junction's state per second -- versus temperature. The frequency increases with temperature and presents a plateau at the frequency of the drive $F_{ac}$. This plateau is a signature of stochastic resonance and noise-induced synchronization~\cite{neiman_stochastic_1998,freund_frequency_2003}.

The junction is considered to be in-phase with the drive if it is in the AP state when the drive is +V and in the P state when the drive is -V. In consequence we can define $P_+$ the probability to phase-lock to the drive and $P_-$ the probability to phase-unlock:

 \begin{align}
 P_+ &= P_{P\rightarrow AP}\left(+V,\frac{T_{ac}}{2}\right)=P_{AP\rightarrow P}\left(-V,\frac{T_{ac}}{2}\right) \nonumber \\
  &= 1-\exp\left(-\frac{T_{ac}}{2}\phi_+\right)
 \end{align}
 \begin{align}
 P_- &= P_{P\rightarrow AP}\left(-V,\frac{T_{ac}}{2}\right)=P_{AP\rightarrow P}\left(+V,\frac{T_{ac}}{2}\right) \nonumber \\
  &= 1-\exp\left(-\frac{T_{ac}}{2}\phi_-\right)
 \end{align}
 where 
 %damien : peut etre plutot reutiliser eqns (3) et (4) ?
 \begin{equation}
 \phi_+=\phi_0~\exp\left(-\frac{\Delta E}{k_BT}\left(1-\frac{V}{V_c}\right)\right)
 \end{equation}
 \begin{equation}
  \phi_-=\phi_0~\exp\left(-\frac{\Delta E}{k_BT}\left(1+\frac{V}{V_c}\right)\right).
 \end{equation}

 For each temperature, the probabilities $P_+$ and $P_-$ are computed analytically and plotted in Fig.\ref{OneSync}a.\\
  
 At low temperatures (below 85K), the frequency of the junction is lower than the drive frequency $F_{ac}$. Indeed the drive is too weak to induce synchronization by itself because its amplitude is sub-threshold ($V=0.75V_c$).\\
 We investigate the conditions leading to the occurrence of synchronization.
  When the junction and the drive are out-of-phase, they have a probability $P_+$ to phase-lock in the next half-period $\frac{T_{ac}}{2}$ (before the next switch of the drive). When the junction and the drive are in-phase, they have a probability $1-P_-$ to stay phase-locked during the next half-period. As the junction is intrinsically stochastic, there cannot be perfect deterministic synchronization. 
  %\textbf{In this study, we choose the arbitrary probability of 0.99 as a lower boundary for phase-locking to be achieved}. Thus,
 We consider the junction to be synchronized to the drive for $P_+>0.99$ and $P_-<0.01$. This corresponds to the temperature range $85\mathrm{K}<T<241\mathrm{K}$ and indeed we observe that it matches the visual boundaries of the plateau where the frequency of the junction is equal to the frequency of the drive. 
  
 Above $T=241\mathrm{K}$, the probability to phase-unlock $P_-$ is higher than 0.01. We observe that the junction regularly slips out of phase.
 These parasitic oscillations -- here called glitches -- raise the frequency of the junction above $F_{ac}$ and destroy synchronization.
 %faire apparaitre dep exp
 
\subsection{Electrical noise induced synchronization \label{syncN}}
 
%In order to use superparamagnetic tunnel junctions in applications, it can be more practical to be able to control its synchronization without having to modify the temperature. 
Synchronization can also be induced by electrical noise added to the drive signal. 
%We consider a junction with an energy barrier and temperature such that $\frac{\Delta E}{k_B T}=30$ and a threshold voltage $V_c=1~\mathrm{V}$. 
We apply a voltage $U(t)=V_{ac}(t)+N(t)$ to the junction. $V_{ac}(t)$ is a square periodic drive of frequency $F_{ac}=1~\mathrm{MHz}$ and amplitude $V=0.75 \mathrm{V}$. $N(t)$ is a white Gaussian noise with standard deviation $\sigma_{noise}$ and cutoff frequency $F_{noise}=100\mathrm{MHz}$. The escape rates $\phi_+$ and $\phi_-$ are time dependent random variables:
  \begin{equation}
 \phi_+(t)=\phi_0~\exp\left(-\frac{\Delta E}{k_BT}\left(1-\frac{V+N(t)}{V_c}\right)\right)
 \end{equation}
 \begin{equation}
  \phi_-(t)=\phi_0~\exp\left(-\frac{\Delta E}{k_BT}\left(1+\frac{V+N(t)}{V_c}\right)\right)
 \end{equation}
 In this case, the switching probabilities need to be averaged over all possible values of $N$. The average probabilities to phase-lock and unlock can be computed as follows:
 \begin{widetext}
 \begin{equation}
\langle P_{\pm}\rangle =1-\Bigg(\int_{-\infty}^{+\infty} \left(1-\exp\left(-\delta t\phi_0\exp\left(-\frac{\Delta E}{kT}\left(1\pm\frac{V+N}{Vc}\right)\right)\right)\right)\psi (N)dN\Bigg)^{\frac{T_{ac}}{2\delta t}},
\end{equation}
 \end{widetext}

where $\delta t=1/F_{noise}$ is the smallest time scale of the electrical noise and $\psi (N)$ is a Gaussian distribution with standard deviation $\sigma_{noise}$.\\
%The numerical simulation is conducted in the same way as in the thermal noise study (subsection \ref{syncT}). 
In Fig.\ref{OneSync}b, the frequency of the junction obtained by numerical simulations as well as the average probabilities $\langle P_+\rangle $ and $\langle P_-\rangle $ are plotted versus the noise's standard deviation $\sigma_{noise}$. The frequency exhibits again a plateau at $F_{ac}$. Noise-induced synchronization with the drive $V_{ac}$ is achieved for an optimal range of electrical noise $0.08V<\sigma_{noise}<0.39V$ which corresponds to $\langle P _+\rangle >0.99$ and $\langle P_-\rangle <0.01$. \\

In conclusion, thermal and electrical noises induce nearly identical behaviors in the single junction situation. For an optimal range of noise, the superparamagnetic tunnel junction is synchronized to the voltage drive.

\section{Noise-induced synchronization of two electrically coupled junctions \label{2sync}}

We now use our model to study the synchronization of coupled oscillators. We consider two identical junctions connected in series. The voltages $V_{1}$ and $V_{2}$ applied to each junction depend on the states of both junctions:
\begin{equation}
V_{1,2}=V_{ac}\frac{R_{1,2}}{R_1+R_2},
\end{equation}
where $R_{1,2}$ are the respective resistances of the junctions.
When the pinned layer of junction 1 is connected to the free layer of junction 2 the junctions are said to be head to tail. On the contrary, when the pinned layer of junction 1 is connected to the pinned layer of junction 2 the junctions are said to be head to head.

\subsection{Thermal noise induced synchronization \label{2syncT}}
\begin{figure*}
 \includegraphics[scale=0.5]{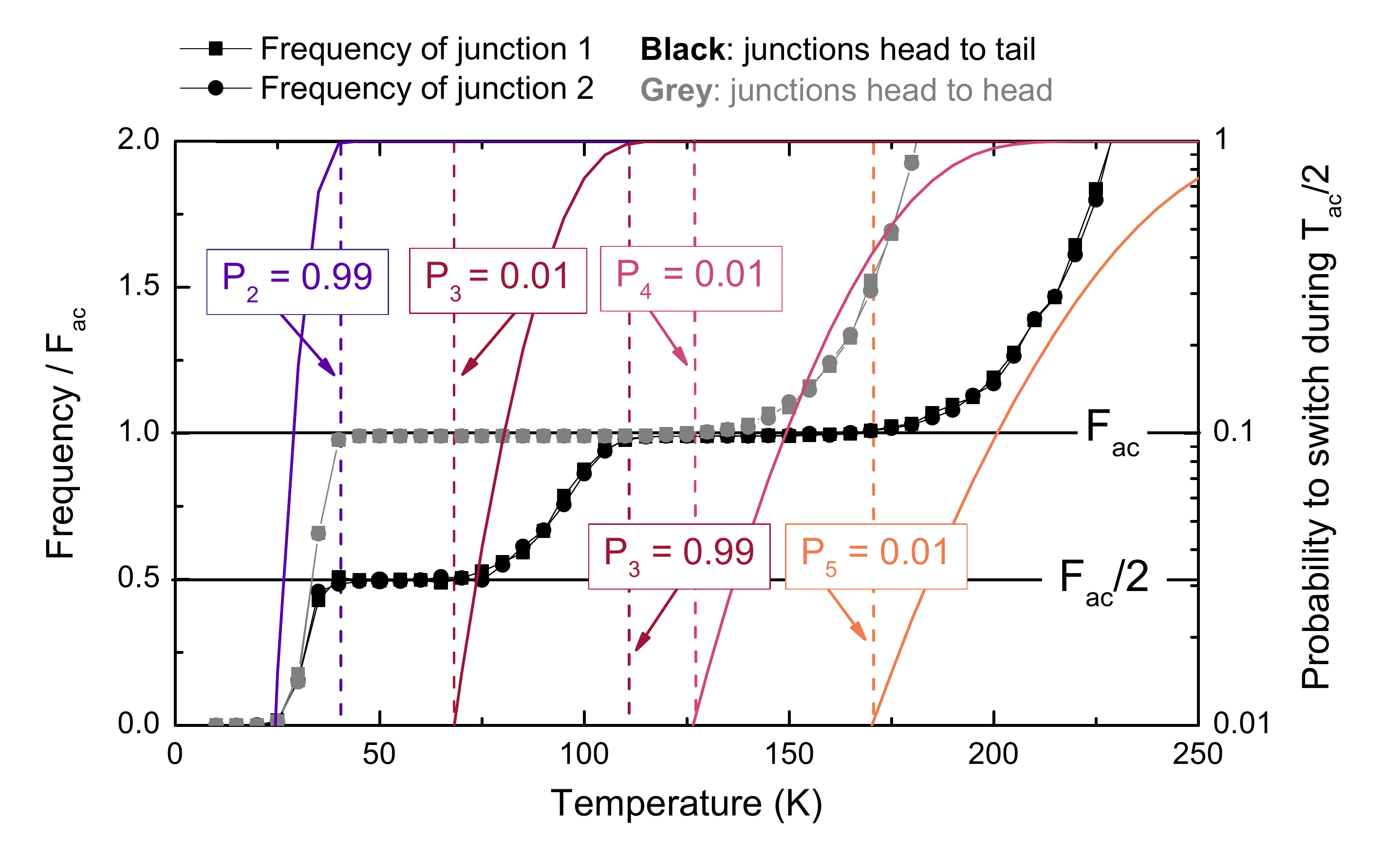}
 \caption{The frequency of junctions is plotted versus temperature. The frequency of junction 1 (resp. 2) is represented with squares (resp. circles). In the case where both junctions are head to tail (resp. head to head) the symbols are black (resp. grey). The drive frequency $F_{ac}$ as well a half the drive frequency $\frac{F_{ac}}{2}$ are represented by black horizontal lines.
The probabilities to switch during $\frac{T_{ac}}{2}$, $P_2$, $P_3$, $P_4$ and $P_5$ are represented by solid lines of color from purple to orange. The dashed vertical lines represent the temperatures at which $P_2 =0.99$, $P_3 = 0.01$, $P_3 = 0.99$, $P_4 = 0.01$ and $P_5 = 0.01$.
\label{TwoSyncTh}}
\end{figure*}

In this section we study the case where only thermal noise is involved. We consider two identical junctions which electrical resistances are such that $R_{AP}=4R_P$, with $R_{AP}$ ($R_P$) the resistance corresponding to the AP (P) state. The corresponding tunnel magneto-resistance is $TMR=\frac{R_{AP}-R_P}{R_P}=300\%$. 
We apply a square periodic voltage of frequency $F_{ac}=10\mathrm{kHz}$ and amplitude $V_{ac}=1.5\mathrm{V}$ to the junctions.\\

\subsubsection{Junctions connected head to tail \label{2syncTs}}

The numerical simulation is conducted in the same way as the single junction study. A random number is generated for each junction. If both junctions have a random number lower than their switching probability, the precise times of switching corresponding to each random number are computed. Two cases arise.
Either the times of switching are separated by less than 1~ns and both junctions switch simultaneously (1~ns is the typical duration of a switch~\cite{sun_spin_2006}). Either the times of switching are separated by more than 1~ns, in which case the junction with the lowest switching time switches first. 
%damien je suggere enlever la phrase qui suit
Then the switching time of the other junction has to be recomputed, because its switching probability is different now that the first junction has switched, modifying the applied voltage.\\

\begin{figure*}
 \includegraphics[scale=0.5]{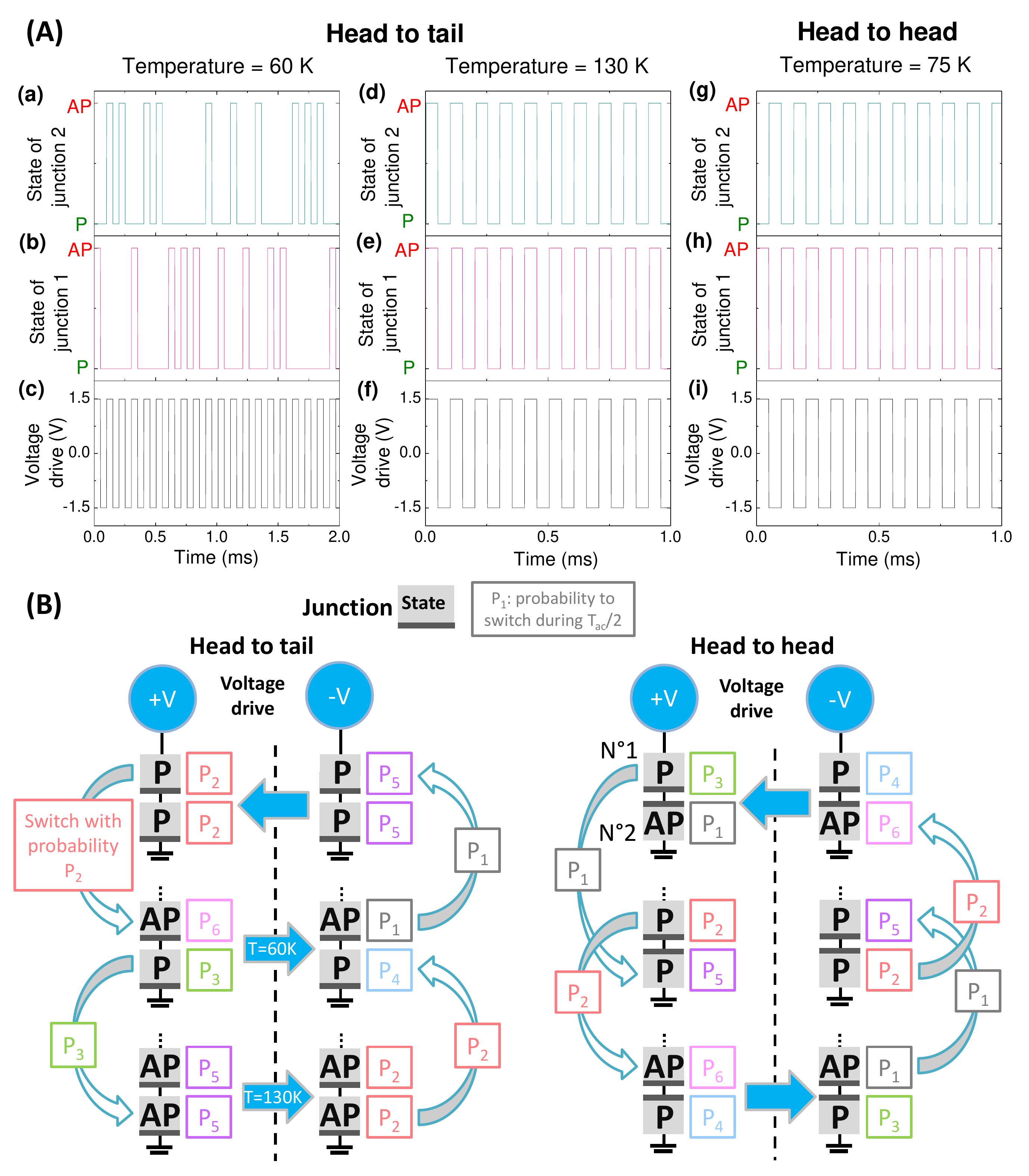}
 \caption{
 A: Panels a, b and c represent the temporal evolution -- at a temperature of $T=60K$ -- of the amplitude of drive voltage $V_{ac}$ (c), the state of junction 1 (b) and the state of junction 2 (a) when the junctions are head to tail. 
Panels d, e and f represent the temporal evolution -- at a temperature of $T=130K$ -- of the amplitude of drive voltage $V_{ac}$ (f), the state of junction 1 (e) and the state of junction 2 (d) when the junctions are head to tail.
Panels g, h and i represent the temporal evolution - at a temperature of $T=150K$ - of the amplitude of drive voltage $V_{ac}$ (g), the state of junction 1 (h) and the state of junction 2 (i) when the junctions are head to head. \\
B: Schematic depiction of the switching cycles that junctions 1 and 2 undergo. The grey rectangles represent the two junctions with their respective state (P or AP). For each configuration, the probability to switch during $\frac{T_{ac}}{2}$ is indicated on the left of each junction. Blue disks indicate the state of the drive voltage ($+V$ or $-V$). Blue horizontal arrows represent switches of the drive while curved vertical arrows represent switches of the junctions. The probability of each switch is indicated on the arrow. When the junctions are head to head, the top junction is 1 and the bottom junction is 2.\label{Schema}}
 \end{figure*}

The frequencies of both junctions are plotted in black in Figure \ref{TwoSyncTh}. They are superimposed for all temperatures. The plateau at the frequency of the drive, characteristic of synchronization, is present between $T=111\mathrm{K}$ and $T=170\mathrm{K}$. Surprisingly, we observe the existence of a second plateau at half the frequency of the drive, between $T=40\mathrm{K}$ and $T=68\mathrm{K}$ . Figure \ref{Schema}A presents
the evolution with time of the voltage drive and the resistances of both junctions. In panels (d), (e) and (f) the temperature is 130K which corresponds to the plateau at the drive frequency. As expected, both junctions are locked in phase with the drive. In panels (a), (b) and (c) the temperature is 60K which corresponds to the plateau at half the drive frequency. We observe that, at each period of the drive, one junction is locked in phase with the drive while the other is blocked in the P state. The phase-locked junction alternates randomly from period to period.\\
In the Appendix, we use further numerical simulations and our analytical model to investigate the influence of the tunnel magneto-resistance and the drive frequency on the existence and width of the plateaus. We observe that a high tunnel magneto-resistance widens the plateau at $F_{ac}/2$ and narrows the plateau at $F_{ac}$. A $300\%$ TMR is a good illustration of the effect of the coupling while maintaining a realistic tunnel magneto-resistance. A drive frequency of $F_{ac}=10kHz$ enables the system to exhibit two significant synchronization plateaus, at $F_{ac}$ and $F_{ac}/2$. 

To get a quantitative understanding of this phenomenon we define six different probabilities which correspond to the possible switching probabilities of the junctions for the different drive and resistances configurations:
\begin{itemize}
\item $P_1=P_{AP\rightarrow P}(-\frac{4V}{5})$
\item $P_2=P_{AP\rightarrow P}(-\frac{V}{2})=P_{P\rightarrow AP}(+\frac{V}{2})$
\item $P_3=P_{P\rightarrow AP}(+\frac{V}{5})$
\item $P_4=P_{P\rightarrow AP}(-\frac{V}{5})$
\item $P_5=P_{AP\rightarrow P}(+\frac{V}{2})=P_{P\rightarrow AP}(-\frac{V}{2})$
\item $P_6=P_{AP\rightarrow P}(+\frac{4V}{5})$
\end{itemize}
When both junctions are in the same state, a voltage $\frac{V_{ac}}{2}$ is applied to each. When they are in different states, $\frac{4V_{ac}}{5}$ is applied to the junction in the AP state while $\frac{V_{ac}}{5}$ is applied to the junction in the P state.

Therefore, $P_1$ and $P_6$ correspond to the switching probabilities of a junction in the AP state when the other junction is in the P state. $P_3$ and $P_4$ correspond to the switching probabilities of a junction in the P state when the other junction is in the AP state. $P_2$ and $P_5$ correspond to the switching probabilities when both junctions are in the same state, P or AP.

In Figure~\ref{TwoSyncTh} the frequencies of junctions 1 and 2 as well as the probabilities $P_2$, $P_3$, $P_4$ and $P_5$ are plotted versus temperature. The probabilities  $P_1$ and $P_6$ do not appear on this graph as $P_1\simeq 1$ and $P_6<0.01$ within the studied temperature range.
Figure~\ref{Schema}B illustrates the switching cycles of both junctions.

The plateau at half the drive frequency can be interpreted as follows and depicted on Fig. \ref{Schema}B. We use the following example as a starting point: the drive voltage is $+V=1.5\mathrm{V}$ and both junctions are in the P state. 
(A similar reasoning can be made for any other initial conditions.) Fig.~\ref{TwoSyncTh} shows that $P_2>0.99$. Therefore, both junctions have a high probability ($P_2$) to switch in the AP state in the following half period $\frac{T_{ac}}{2}$. One junction switches to the AP state as depicted in Fig.~\ref{Schema}B. The switching probabilities are $P_6$ for the junction in the AP state and $P_3$ for the junction in the P state. As both probabilities $P_6$ and $P_3$ are below 0.01, both junctions remain in their state until the next switch of the drive, depicted by the "$T=60\mathrm{K}$" arrow in Fig.~\ref{Schema}A. When the drive switches from $+V$ to $-V$ the switching probabilities become $P_1$ for the junction in the AP state and $P_4$ for the junction in the P state. As $P_4<0.01$ while $P_1>0.99$, the junction in the AP state switches to the P state. Both junctions are in the P state with a switching probability of $P_5<0.01$: they remain in their state until the next switch of the drive. Then, their switching probabilities become $P_2$ and the cycle starts again. To summarize, at each oscillation cycle of the drive, one random junction is locked in phase with the drive while the other is blocked in the P state, leading to a mean frequency of $\frac{F_{ac}}{2}$ for both junctions. This is depicted on the the "Temperature = 60K" panel of Fig.~\ref{Schema}B.

For $111\mathrm{K}<T<170\mathrm{K}$, $P_3$ is larger than 0.99. Considerations similar the the study of the $40 - 68\mathrm{K}$ range, depicted in  Fig.~\ref{Schema}A, show that both junctions are locked in phase with the drive. This gives rise to the frequency plateau at $F_{ac}$ and is illustrated on the "Temperature = 130K" panel of Fig.~\ref{Schema}B.
%the cycle starts again with the switch of one of the junctions from the P to the AP state. The switching probabilities are $P_6$ for the junction in the AP state and $P_3$ for the junction in the P state. $P_6$ is still below 0.01, but now $P_3>0.99$. Therefore, the junction in the P state will switch in the AP state as well. Both junction now have a switching probability of $P_5<0.01$ therefore they remain in the AP state until the next switch of the drive, depicted by the "$T=130\mathrm{K}$" arrow on Fig.~\ref{Schema}A. Both junctions then have a switching probability of $P_2>0.99$ therefore one of the junctions will switch in the P state. The junction remaining in the AP state has a switching probability $P_1>0.99$ so it switches in the P state as well. Both junctions then have a $P_5<0.01$ switching probability so they remain in the P state until the drive switches and the cycle starts again. 
%To summarize, both junctions are phase-locked to the drive - as depicted on the "Temperature = 130K" panel of Fig.~\ref{Schema}B - and thus have a mean frequency of $F_{ac}$.

%The $68\mathrm{K}$ to $111\mathrm{K}$ temperature range corresponds to an intermediate regime where $0.01<P_3<0.99$. In consequence the two switching cycles are possible. As the temperature increases, $P_3$ increases and the oscillations where both junctions are phase-locked to the drive are more and more probable. 

$P_5$ is the probability to phase-unlock with the drive, both when the drive is in the $+V$ and $-V$ states (Fig.\ref{Schema}A). For temperatures above 170K, $P_5$ is larger than 0.01 so glitches appear and synchronization is destroyed.

\subsubsection{Junctions connected head to head \label{2syncTo}}
We now consider that junction 2 is connected head to head with junction 1. 
In that case, the voltage applied to the junction, and governing the probabilities is $V_2=-V_{ac}\frac{R_2}{R_1+R_2}$. We observe in Fig. \ref{TwoSyncTh}(a) that the frequencies of junctions 1 and 2 -- plotted in grey -- are equal at all temperatures. The frequency of the junctions increases with the temperature. Contrary to the case where the junctions are head to tail, a single plateau at the drive frequency $F_{ac}$ is observed, between $T=40K$ and $T=126K$. Panels (g), (h) and (i) of Fig.~\ref{Schema}B depict the evolution with time of the voltage drive, and the states of junctions 1 and 2, for a temperature of T=75K. This regime corresponds to the plateau and we observe that junction 1 is locked in phase with the drive while junction 2 is locked in anti-phase with the drive. These results can be interpreted with the same reasoning as in the head to tail case and as depicted in Fig.~\ref{Schema}A.\\

\subsection{Electrical noise induced synchronization \label{2syncN}}

\begin{figure*}
 \includegraphics[scale=0.5]{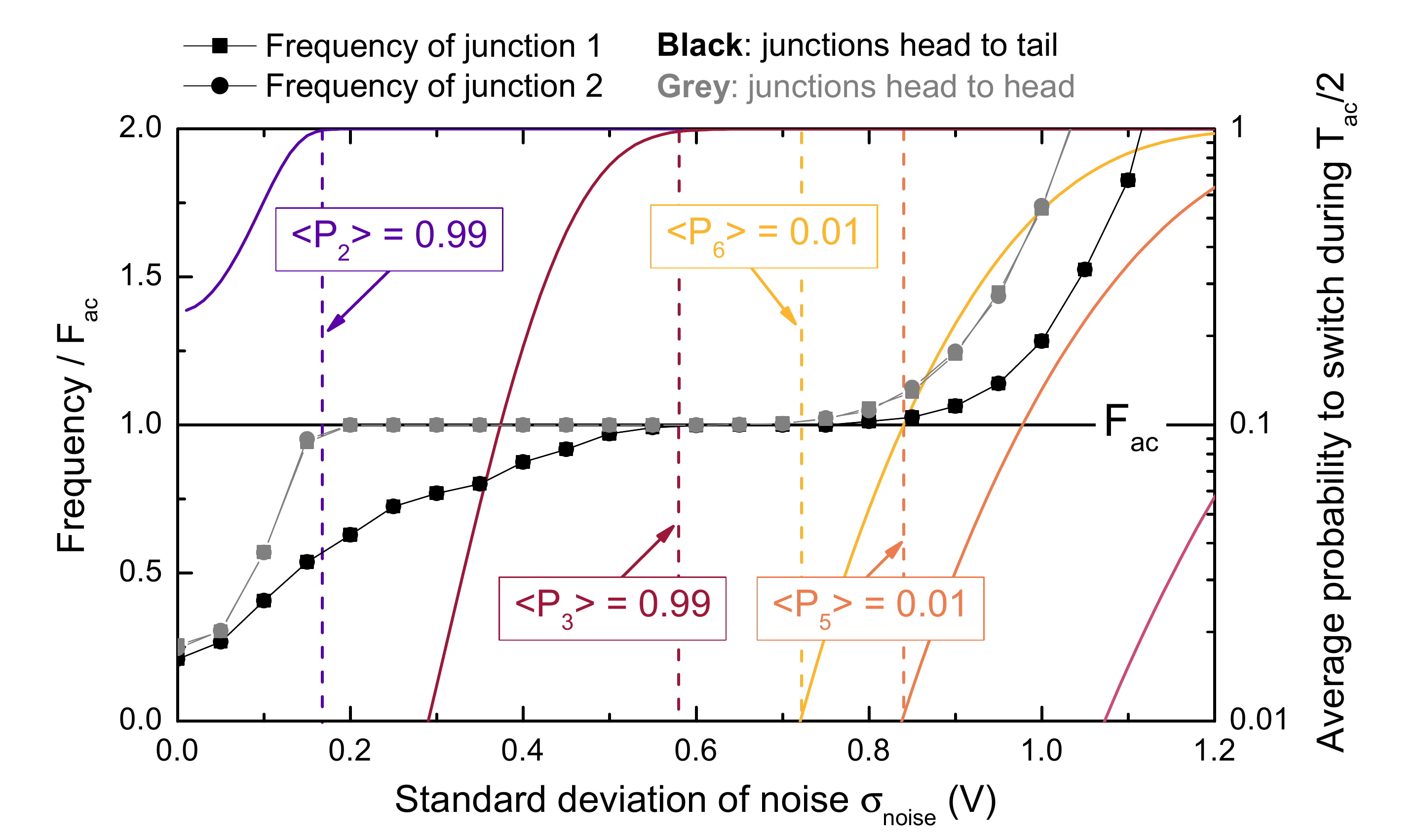}
 \caption{The evolution of the frequency of the junctions is plotted versus the standard deviation of the Gaussian electrical noise. The frequency of junction 1 is represented with squares and the frequency of junction 2 is represented with circles. In the case where both junctions are head to tail the symbols are black whereas in the case where the junctions are head to head the symbols are grey. The drive frequency $F_{ac}$ as well a half the drive frequency $\frac{F_{ac}}{2}$ are represented by black horizontal lines.
The average probabilities to switch during $\frac{T_{ac}}{2}$, $P_1$, $P_2$, $P_3$, $P_4$, $P_5$ and $P_6$ are represented by solid lines of color from purple to light orange. The dashed vertical lines represent the electrical noise levels at which $P_2 =0.99$, $P_3 = 0.99$, $P_5 = 0.01$ and $P_6 = 0.01$.
\label{TwoSyncElec}}
\end{figure*}

In this section, we now control synchronization by injection of electrical noise. Contrary to the single junction case, we show that, in the two junctions case, electrical noise induces a behavior qualitatively different from the one induced by thermal noise. We consider two identical junctions which a tunnel magneto-resistance of $TMR=\frac{R_{AP}-R_P}{R_P}=100\%$. As shown in the Appendix, this is high enough to have a signal that can be detected easily and low enough to exhibit a significant synchronization plateau.\\
%The junctions are submitted to a voltage $U(t)=V_{ac}(t)+N(t)$. $V_{ac}$ is a square periodic drive of frequency $F_{ac}=1~\mathrm{MHz}$ and amplitude $1.5\mathrm{V}$. $N$ is a white Gaussian noise of standard deviation $\sigma_{noise}$ and cutoff frequency $F_{noise}=100~\mathrm{MHz}$. 
The switching probabilities are the averaged probabilities $\langle P_1\rangle $ to $\langle P_6\rangle $ over all possible values of $N(t)$.  \\
%The numerical simulations are conducted in the same way as in section \ref{2syncT}.
In Fig.\ref{TwoSyncElec} the frequencies of junctions 1 and 2 as well as the average probabilities $\langle P_1\rangle $ to $\langle P_6\rangle $ are plotted versus the standard deviation of the noise distribution, both for the configurations where the junctions are head to tail and head to head. 

\subsubsection{Junctions connected head to tail\label{2syncNs}}
Both junctions have the same frequency at all levels of noise. We do not observe any synchronization at $\frac{F_{ac}}{2}$ for the junctions frequency in the range of noise for which $\langle P_2\rangle >0.99$ and $\langle P_3\rangle <0.01$. This can be interpreted as follows. In the thermal noise study, the escape rates are constant in time for a given temperature. The probability of switching at a given time is low but, because it is sustained over a long time, the probability for one junction to switch eventually is high. The situation is different when electrical noise is added on the drive. The distribution of the voltage from the electrical noise is Gaussian. 
As the noise is on the drive which is applied to both junctions, when a high value voltage occur, both junctions have a probability to switch that is close to one. Thus both junctions switch simultaneously. Therefore, when the drive switches to $+V$, either a high value voltage event occurs and both junctions switch simultaneously in the AP state;
either no high voltage event occur, and only one junction switches in the AP state while the other remains blocked in the P state as explained in section \ref{2syncTs}. As the noise level is increased, more high value voltage events occur and thus more simultaneous switches of the junctions occur. These observations pin a fundamental physical difference between internal and external noise. \\
  \\ 
For $0.58\mathrm{V}<\sigma_{noise}<0.84\mathrm{V}$ the frequency of the junctions is equal to the drive frequency $F_{ac}$. Indeed $\langle P_3\rangle >0.99$ and $\langle P _5\rangle <0.01$. Both junctions are synchronized with the drive, in the same way as in the thermal noise study. For $\sigma_{noise}>0.84\mathrm{V}$, $\langle P _5\rangle$ is larger than $0.01$ so glitches appear and synchronization is lost.

\subsubsection{Junctions connected head to head\label{2syncNo}}
Both junctions have the same frequency for all levels of noise. For $0.16\mathrm{V}<\sigma_{noise}<0.72\mathrm{V}$ the frequency of the junctions is equal to the frequency of the drive $F_{ac}$. 
The behaviors of both junctions are the same as in the thermal noise study: junction 1 is locked in-phase while junction 2 is locked in anti-phase with the drive.  $\langle P_6\rangle > \langle P_4\rangle $ so synchronization is lost when $\langle P_6\rangle >0.01$, which occurs when $\sigma_{noise}>0.72\mathrm{V}$.

\section{Conclusion}
In this work we have investigated the noise-induced synchronization of one then two electrically coupled stochastic magnetic oscillators with a periodic drive.  We have shown that both thermal and electrical noise can induce synchronization of two coupled superparamagnetic tunnel junctions with a sub-threshold voltage drive. When the junctions are head to tail, both are synchronized in-phase with the drive for an optimal level of noise. When the junctions are head to head, one is synchronized in-phase with the drive while the other is synchronized in anti-phase. Furthermore, this study has enabled us to highlight a fundamental difference between the synchronization phenomena induced by thermal and electrical noise. Indeed, the fact that electrical noise makes the escape rates probabilistic themselves and that the electrical noise is correlated for both junctions lead to simultaneous switches of the two junctions - while this does not occur for thermal noise. We have provided an intuitive analytic understanding of how one or several coupled stochastic oscillators respond to a periodic drive.

This model will enable investigating the interactions and synchronization of several stochastic oscillators in various configurations. The results will be useful for understanding and designing networks of coupled stochastic oscillators. Among potential applications of such networks, bio-inspired computing is particularly promising, in particular as it allows taking advantage of noise and stochasticity for processing information. Indeed networks of coupled oscillators can perform pattern recognition and classification through synchronization \cite{hoppensteadt_oscillatory_1999,aonishi_phase_1998}. Implementing these networks with sub-threshold synchronization of superparamagnetic tunnel junctions would allow achieving such cognitive tasks at low energy cost.

\appendix*
\section{Effects of magneto-resistance and drive frequency}
\begin{figure*}
 \includegraphics[scale=0.5]{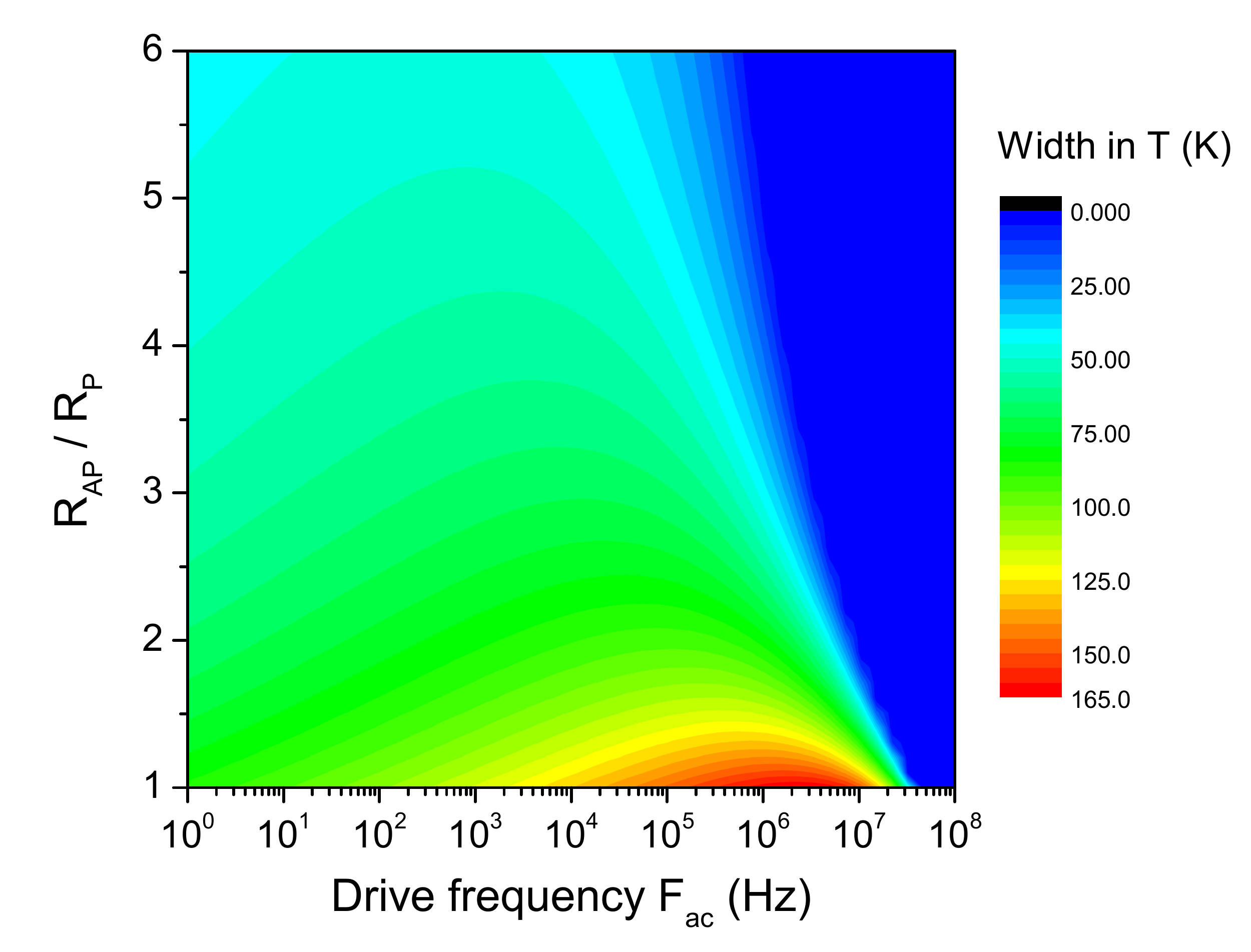}
 \caption{Width in temperature of the synchronization plateau  $F=F_{ac}$ , in function of the $R_{AP}/R_P$ ratio and the drive frequency. 
\label{map2}}
\end{figure*}

\begin{figure*}
 \includegraphics[scale=0.5]{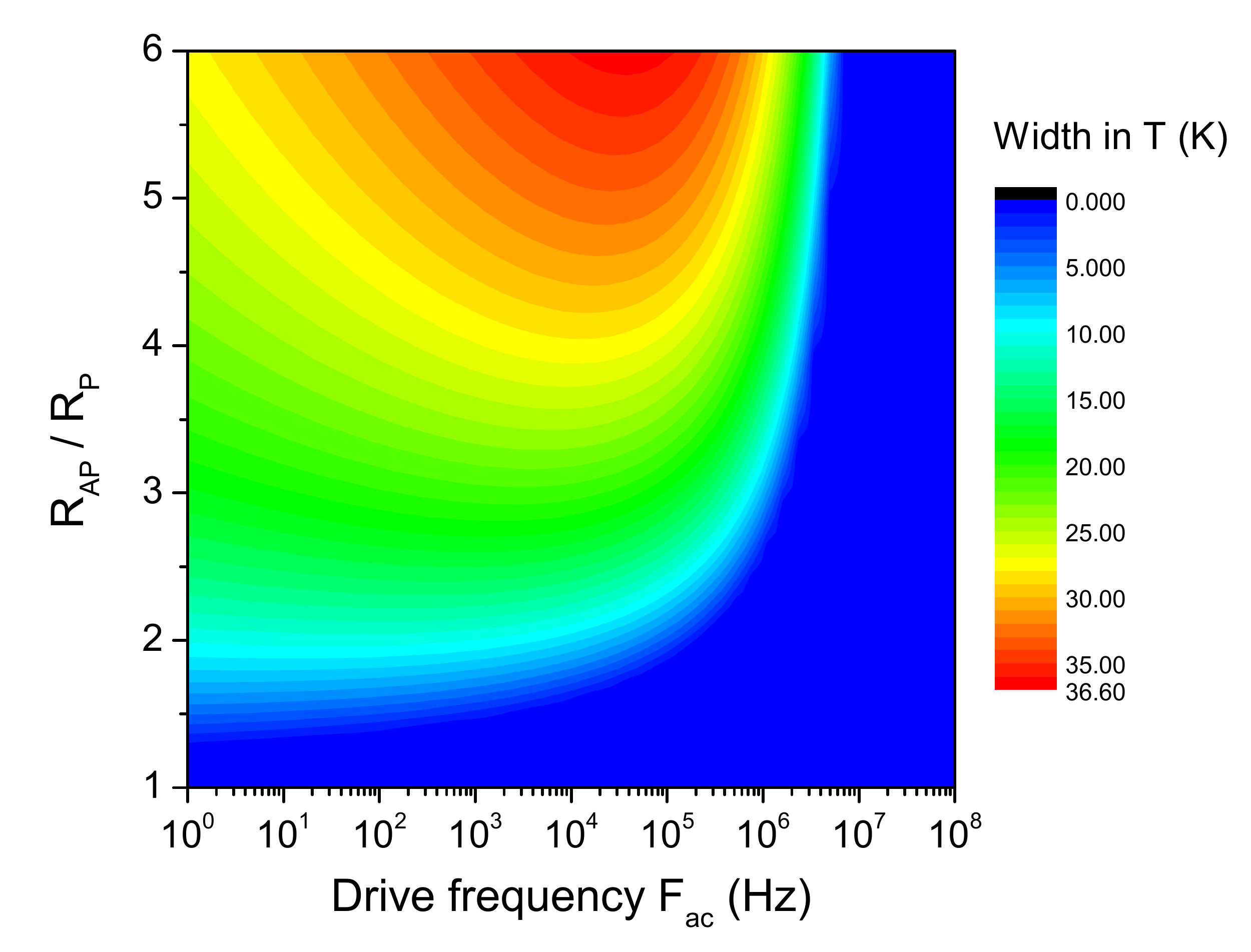}
 \caption{Width in temperature of the synchronization plateau  $F=F_{ac}/2$, in function of the $R_{AP}/R_P$ ratio and the drive frequency. 
\label{map2_half}}
\end{figure*}

In this appendix we study the effects of tunnel magneto-resistance and drive frequency on noise induced synchronization of two superparamagnetic tunnel junctions, connected head to tail. 
Two identical superparamagnetic tunnel junctions of energy barrier such that $\Delta E=30k_BT$ studied at $T=50K$ are submitted to a periodic drive of amplitude $V=1.5V_c$. The analytical model is used to compute the width in temperature of  both synchronization plateaus, for various drive frequencies and magneto-resistance values. The results are presented in Figure~\ref{map2} for the $F=F_{ac}$ plateau and in Figure~\ref{map2_half} for the $F=F_{ac}/2$ plateau.

We observe that increasing the tunnel magneto-resistance widens the plateau at $F=F_{ac}/2$ and narrows the plateau at $F=F_{ac}$. When the junctions are in different states, the ratio between the voltage received by the junction in the AP state and the junction in the P state increases with the $R_{AP}/R_P$ ratio, which strengthens the coupling and favors the $F=F_{ac}/2$ regime.  For $R_{AP}=R_P$ there is no tunnel magneto-resistance and the plateau at $F=F_{ac}/2$ is absent. The effect of drive frequency is non-monotonous. Drive frequencies too high compared to the natural frequency of the junctions (which depends on the energy barrier) prevent the apparition of synchronization plateaus. On the other side, as temperature increases the frequency of the junction,  lowering the drive frequency narrows the temperature range which can induce synchronization. 

\begin{figure*}
 \includegraphics[scale=0.5]{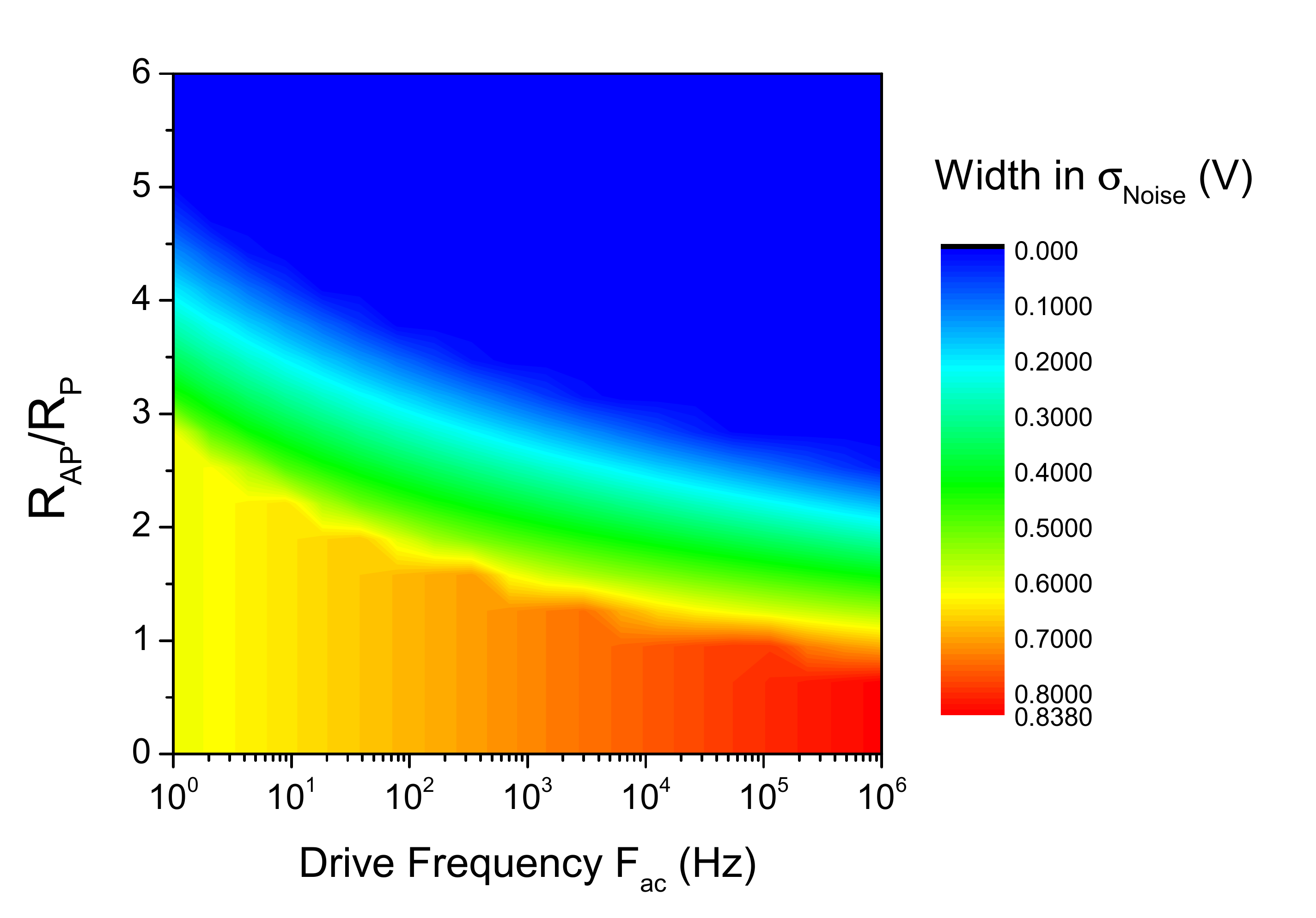}
 \caption{Width in noise of the synchronization plateau  $F=F_{ac}$, in function of the $R_{AP}/R_P$ ratio and the drive frequency. 
\label{map_noise}}
\end{figure*}

The same study is conducted in the case of synchronization induced by electrical noise. Figure \ref{map_noise} presents the influence of drive frequency and tunnel magneto-resistance on the width of the synchronization plateau at $F_{ac}$. We observe that the plateau is widest for low magneto-resistances and high drive frequencies. Numerical simulations show that no synchronization plateau can be observed at $F_{ac}/2$, no matter the value of the magneto-resistance.\\

\begin{acknowledgments}
The authors acknowledge financial support from the FET-OPEN Bambi project No. 618024 and the ANR MEMOS No. ANR-14-CE26-0021-01. A. M. acknowledges financial support from the Ile-de-France regional government through the DIM Nano-K program. N. L. acknowledges financial support from the French National Research Agency (ANR) as part of the Investissements d’Avenir program (Labex NanoSaclay, reference: ANR-10-LABX-0035)
\end{acknowledgments}

\bibliography{Zotero}

\end{document}